\documentclass[aps,amsmath,amssymb,twocolumn,groupedaddress]{revtex4-1}  
\usepackage{graphicx}
\usepackage{hyperref}
\usepackage{natbib}
\usepackage{xcolor}
\usepackage{braket}

\definecolor{dbl}{rgb}{0, 0, 0.9}
\hypersetup{
  bookmarksopen=true,     
  colorlinks=true,        
  linkcolor=dbl,     
  citecolor=dbl,      
  filecolor=dbl,      
  urlcolor=dbl        
}

\definecolor{myblue}{rgb}{0.05,0.1,0.5}

\definecolor{myred}{rgb}{0.5,0.05,0.1}

\begin{document}
\title{Displacement of ultra-high-energy cosmic ray source images by the intergalactic magnetic field: the cases of Cen A and M83.}

\author{K.~Dolgikh$^{1,2}$, A.~Korochkin$^{3}$,  G.~Rubtsov$^{1}$, D.~Semikoz$^{4}$, and I.~Tkachev$^{1,5}$}
\affiliation{$^{1}$ Institute for Nuclear Research of the Russian Academy of Sciences, Moscow, 117312 Russia}
\affiliation{$^{2}$ Lomonosov Moscow State University, Faculty of Physics Moscow, 119991 Russia} 
\affiliation{$^{3}$ Université Libre de Bruxelles, CP225 Boulevard du Triomphe, 1050 Brussels, Belgium}
\affiliation{$^{4}$ APC, Universit\'e Paris Cit\'e, CNRS/IN2P3, CEA/IRFU, Observatoire de Paris, 119 75205 Paris, France}
\affiliation{$^{5}$ Novosibirsk State University, Novosibirsk 630090, Russia}

\begin{abstract}
The standard assumption about the influence of the turbulent intergalactic magnetic field (IGMF) on the images of ultra-high-energy cosmic rays (UHECR) sources is that the latter are formed in a random walk mode in the deflection angle. As a result, the images are symmetrically broadened to angular scales proportional to the IGMF strength and the square root of its correlation length.
We demonstrate that when the size of the emitting region is smaller than the correlation length of the IGMF, a new focusing regime emerges. In this regime, significant deviations from the standard random walk approximation occur even when the distance between the source and the observer exceeds several tens of IGMF correlation lengths. The angular size of the resulting images is typically smaller than predicted by random walk, and the IGMF causes a systematic shift of the entire image away from the true source direction.
This introduces additional uncertainty in the search for UHECR sources. We show that the excess observed by Pierre Auger Observatory in the direction of Cen~A can be explained by the contribution of M83, provided that the image shift occurs as predicted by some models of the Galactic magnetic field (GMF) and that the IGMF plays a minor role due to its low strength and short coherence length. Alternatively, Cen~A may indeed be the true source of the excess, as certain realizations of the IGMF can compensate the deflection caused by the GMF.
\end{abstract}

\maketitle

\section{Introduction}
Despite significant effort, the sources of ultra-high-energy cosmic rays (UHECRs) remain unknown. The sky at the highest energies, as observed by the two largest modern UHECR observatories, the Pierre Auger Observatory (Auger) \cite{PierreAuger:2015eyc} and the Telescope Array (TA) \cite{2012NIMPA.676...54T, 2012NIMPA.689...87A} --- appears to be remarkably isotropic. Even the high statistics of events collected above $E>10^{19}$~eV by Auger and TA do not allow for an unambiguous identification of any specific class of UHECR accelerators. This makes the slight deviations from isotropy even more valuable and may provide hints about the UHECR sources.

At the largest angular scales, Auger has detected a dipole in UHECR arrival directions, whose significance has surpassed $5\sigma$ for energies $E > 8$~EeV \cite{2017Sci...357.1266P, 2024ApJ...976...48A}. Its presence has been confirmed using combined data from Auger and TA \cite{2022icrc.confE.375T, PierreAuger:2023mvf}. Additionally, at intermediate angular scales, Auger and TA observe several hotspots of approximately $20^\circ$ in size. The most significant excess in UHECR flux is in the direction of the Centaurus region, detected by Auger at a significance level of $4\sigma$ at the energies $E>40$~EeV \cite{PierreAuger:2022axr,PierreAuger:2023fcr}. The TA hotspot located near the Ursa Major region has a significance level of $3.4\sigma$ sigma at similar energies~\cite{TelescopeArray:2014tsd}. In addition, there are indications of an additional hotspot near the Perseus-Pisces supercluster~\cite{2021arXiv211014827T,Kim:2025ykm}. 

Comparison of hotspot positions with source catalogs reveals a correlation with nearby starburst galaxies~\cite{PierreAuger:2018qvk, 2023EPJWC.28303002D}. While this correlation has already reached the significance of $4\sigma$, the analysis presented in \cite{PierreAuger:2018qvk, 2023EPJWC.28303002D} does not account for UHECR deflections in the Galactic magnetic field (GMF). However, since the mass composition of the UHECR at energies above $E > 10^{19}$~eV is dominated by intermediate-mass nuclei, these deflections are expected to be strong in a significant portion of the sky, as predicted by the latest coherent GMF models, \texttt{UF23}~\cite{2024ApJ...970...95U} and \texttt{KST24}~\cite{2025A&A...693A.284K}. This can shift the apparent positions of the sources. Therefore, the model describing the correlations must also include deviations in the GMF. On the other hand, uncertainties in GMF modeling remain large, as discussed in detail in \cite{2025arXiv250116158K}, in particular in the application to the observed hotspots.

Apart from the GMF, UHECR propagation can also be affected by intergalactic magnetic fields (IGMF), provided that its strength is close to the current upper limits of approximately $B \sim 0.1 - 1$~nG. These upper limits obtained from Faraday rotation measures of extragalactic sources are valid for the volume-filling magnetic field, which resides in the voids of the large-scale structure (LSS)~\cite{2016PhRvL.116s1302P, 2024arXiv241214825N}. As a result, they do not directly constrain the magnetic field in the local Universe, where it could be stronger. This could naturally occur if the Milky Way resides in a filament of the LSS~{\cite{2021MNRAS.505.4178V}}, or if the local magnetic field has been enhanced by feedback from, for example, the Council of Giants, a group of nearby large galaxies{~\cite{2014MNRAS.440..405M,Taylor:2023qdy}}.

The impact of IGMF on the interpretation of UHECR data has been extensively studied in the literature. At the highest energies, when the distance to the source is much larger than the IGMF correlation length $D \gg L_\mathrm{c}$, UHECR propagation in the IGMF follows a random walk in the deflection angle. The simplest and computationally efficient way to account for this effect is the smearing angle approximation, which was adopted, for example, in \cite{2021JCAP...04..065K, 2024PhRvL.133d1001A, 2024PhRvD.110b2006A, 2024ApJ...966...71B}. In this approximation, the initial point-like image of the source is replaced by a symmetric halo with angular size $\delta$. The size $\delta$, in turn, depends on the IGMF strength $B$, the correlation length $L_\mathrm{c}$, and the distance to the source $D$ such that $\delta \propto B\sqrt{D L_\mathrm{c}}/E$. At the same time, the position of the center of the source's image remains unchanged and points to the direction to the source.

The isotropic smearing approximation is valid when $D \gg L_\mathrm{c}$. In contrast, when $D \ll L_\mathrm{c}$, the propagation of UHECRs is ballistic, and the source image remains nearly point-like for a fixed energy. There exists an intermediate regime characterized by $L_\mathrm{c} \approx (0.001 - 1) \times D$ and $R_\mathrm{L} > L_\mathrm{c}$, where $R_\mathrm{L}$ is the particles' Larmor radius. We will refer to this as the focusing regime. This regime has been qualitatively discussed in \cite{1996ApJ...472L..89W, 2016PhRvD..93f3002H}. 

In this work, we continue our study of UHECR propagation in the focusing regime by conducting 3D numerical simulations of UHECR propagation from point sources. In our previous paper, we demonstrated that in this regime, the observed source flux can be amplified or de-amplified by several orders of magnitude compared to the expectation from the inverse-square law \cite{dolgikh2022causticlike}, and its image may be distorted \cite{DOLGIKH20245295}. A distinguishing feature of this intermediate regime is that the particles have not yet forgotten their initial conditions. Particles emitted from a point source in nearby directions propagate through the same magnetic field, and thus their trajectories remain correlated over a sufficiently large distance $>L_\mathrm{c}$, leading to the effects described above.

This paper focuses on the appearance of the source images, logically building upon our previous findings~\cite{dolgikh2022causticlike, DOLGIKH20245295}. We show that in this regime the IGMF not only smears the source image but also shifts its position thus mimicking the effect of the coherent GMF. This complicates the search for UHECR sources by introducing additional uncertainty related to the shift of the source image in the IGMF. Unlike the image shift in the coherent GMF, which is determined by its global configuration, the shift in the IGMF depends on a specific realization of the turbulent field between the source and the observer, and thus is much more difficult to control.

In Section \ref{sec:simulations} we present the results of the paper on the source images in the turbulent IGMF with the Kolmogorov spectrum. In Section \ref{sec:approximation_uncorr} we consider the approximation of uncorrelated trajectories and reproduce analytical results for source image size in this approximation. We show that the results of numerical simulations performed under the condition that each particle passes through its own random field are consistent with analytical calculations. In Section~\ref{sec:image_shift}, we numerically study the source shift effect and explore its dependence on the IGMF parameters and the distance to the source. Next, in Section~\ref{sec:cen_a}, we apply this effect to the Centaurus region excess in the UHECR data and show its interplay with the coherent GMF. We discuss how this effect may influence popular excess interpretation as an image of the nearby starburst galaxies NGC~4945 and M83 or the radio galaxy Cen~A. This is followed by concluding remarks in Section~\ref{sec:conclusions}.


\section{UHECR source images in the focusing regime}
\label{sec:simulations}
Although the propagation of UHECRs is generally determined by the full three-dimensional IGMF structure,
we work here in the approximation when UHECRs escape from their source regions to the regions with relatively weak (compared to the magnetic field around the sources) turbulent IGMF with $B \sim 1$~nG amplitude, which we assume for simplicity to be completely uniform between the observer and the source. 

To construct images of UHECR sources as they would appear after propagation through the IGMF, we performed a series of numerical simulations. For 3D modeling of UHECR propagation we used the publicly available Monte Carlo code \texttt{CRbeam}~\footnote{\protect\url{https://github.com/okolo/mcray}}\cite{Kalashev_2023}. The IGMF was assumed to be spatially uniform with a Kolmogorov spectrum $P_B(k)$ over a wavelength range from $L_\mathrm{max}$ down to $L_\mathrm{min} = L_\mathrm{max}/100$:
\begin{equation}\label{eq:igmf_spec}
    P_B(k) = 
    \begin{cases}
        0 \qquad\qquad\qquad\, k < k_\mathrm{min}\\
        A\left(\frac{k_\mathrm{min}}{k}\right)^{m} \qquad k_\mathrm{min} \le k \le k_\mathrm{max}\\
        0 \qquad\qquad\qquad\, k_\mathrm{max} < k
    \end{cases}
\end{equation}
where $k_\mathrm{min} = 2\pi/L_\mathrm{max}$, $k_\mathrm{max} = 2\pi/L_\mathrm{min}$ and $A$ is the normalization constant, determined by the condition that the root-mean-square (rms) value of the magnetic field equals $B$. The exponent was fixed to $m=11/3$, as for Kolmogorov turbulence. It was generated following the method of~\cite{Giacalone_1999}, with the number of modes of the order of $10^3$, distributed logarithmically in wavelength. The resulting magnetic field is divergence-free. A specific random realization of the IGMF is determined by a random seed value. 

For consistency with previous studies, we present our results in terms of the IGMF correlation length $L_\mathrm{c}$ . We define $L_\mathrm{c}$ following \cite{Harari:2002dy} as
\begin{equation}\label{L_C}
    L_\mathrm{c}=\frac{1}{2} L_\mathrm{max}\frac{n-1}{n}\frac{1-(L_\mathrm{min}/L_\mathrm{max})^n}{1-(L_\mathrm{min}/L_\mathrm{max})^{n-1}}.
\end{equation}
In the case of Kolmogorov turbulence considered in this paper, where $n=5/3$ this yields $L_\mathrm{c} \approx L_\mathrm{max}/5$ for $L_\mathrm{min}\ll L_\mathrm{max}$.  
\begin{figure}
    \begin{minipage}[t]{0.483\textwidth}
        \includegraphics[width=1.07\linewidth]{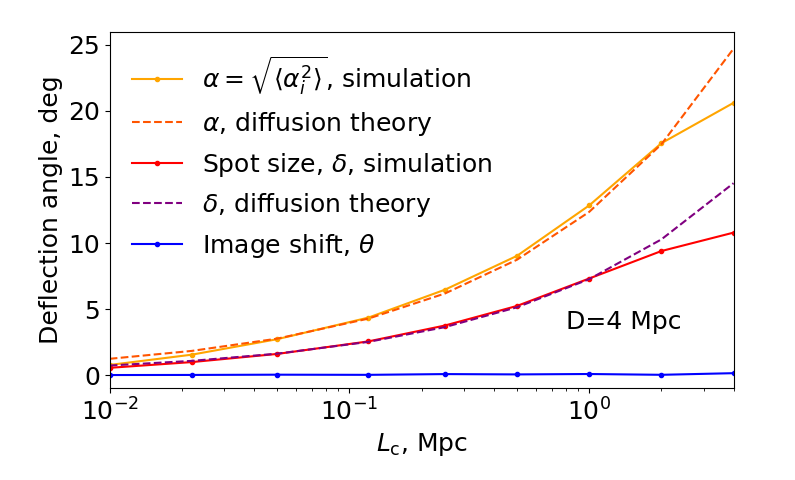}
        \caption{Particles' deflection angle $\alpha$, image size $\delta$ and image shift $\theta$ as a function of the IGMF correlation length for the case of uncorrelated trajectories (each particle was propagated in its own realization of the IGMF, see Sec.\ref{sec:approximation_uncorr} for details). Solid lines represent the RMS values inferred from the numerical simulations using Eq.~(\ref{eq:angles}). Dashed lines show the analytical expectations based on Eq.~(\ref{eq:angles_theor}). The IGMF strength was fixed to $B=1.5$~nG while the distance to the source was set to $D=4$~Mpc.}
        \label{fig:AvgRandomSpotSize}
    \end{minipage}
\end{figure}

We assumed that the source emits UHECRs uniformly within a cone of a given opening angle and axis direction, pointing toward the observer. The opening angle was chosen to be sufficiently large to ensure that the results are independent of its specific value. UHECR energy losses were neglected since in all simulations the distance between the source and the observer does not exceed $D = 10$~Mpc. The rigidity was fixed to $R=10$~EV. To accelerate computations, we used { the idea of} the UHECR aiming method developed in our previous paper \cite{DOLGIKH20245295}.
{This method allows adjusting the parameters of the emission cone (opening angle and axis) based on previous simulations within the same IGMF realization. To further speed up the calculations, we used multiprocessing with \texttt{OpenMP}, which enables the use of multiple CPU cores on a single computer system.}
We consider only steady-state sources, assuming that the source duty cycle is much longer than the UHECR time delay in IGMF and GMF.

Each simulation stops when $N=1000$ particles hit the spherical observer whose radius was set to $R_\mathrm{obs} = 100$~kpc. The value of $N$ was chosen to ensure sufficient event statistics. Before the obtained set of events can be used for further analysis, a correction must be applied to account for the finite size of the observer. The details of this correction were described in \cite{DOLGIKH20245295}, but its essence can be illustrated using the case of straight-line propagation of photons. After the correction, the image of a photon source on a finite-sized observer should appear point-like.

Given the set of detected events emitted with the initial momenta $\mathbf{p}_{i0}$, the image of the source at the observer is determined by their final momenta, $\mathbf{p}_{i}$, where the index $i$ labels individual particles. For subsequent calculations, it is useful to introduce dimensionless unit vectors representing the momentum directions, $\mathbf{n}_{i0} = \mathbf{p}_{i0}/p_{i0}$ and $\mathbf{n}_{i} = \mathbf{p}_{i}/p_{i}$, and a unit vector $\mathbf{s}$ pointing from the source to the observer. The mean final direction of the particles, $\mathbf{n}_\mathrm{avg}$, is then given by
\begin{equation}
    \mathbf{n}_\mathrm{avg} = \frac{\langle\mathbf{n}_i\rangle}{|\langle\mathbf{n}_i\rangle|},
\end{equation}
where the angle brackets denote averaging over the set of events $\langle\mathbf{n}_i\rangle = \frac{1}{N}\sum_{i=1}^{N}\mathbf{n}_i$. The deflection angle $\alpha_i$ of a particle is defined as
\begin{equation}
    \cos\alpha_i = \mathbf{n}_{i}\cdot\mathbf{n}_{i0},
\end{equation}
while the angle $\delta_i$ between a particle's final direction and the mean final direction is given by
\begin{equation}\label{eq:im_size}
    \cos\delta_i = \mathbf{n}_{i}\cdot\mathbf{n}_\mathrm{avg}.
\end{equation}
Averaging over the set of detected particles, we arrive at the following three angles
\begin{equation}\label{eq:angles}
    \begin{cases}
            \alpha = \alpha_\mathrm{rms} = \sqrt{\langle\alpha_i^2\rangle} \\ 
            \delta = \delta_\mathrm{rms} = \sqrt{\langle\delta_i^2\rangle} \\
            \theta = \arccos(\mathbf{n}_\mathrm{avg}\cdot\mathbf{s)}
    \end{cases}
\end{equation}
each with a clear physical meaning. The angle $\alpha$ corresponds to the RMS deflection angle of the particle from its initial direction, $\delta$ represents the angular size of the source's image as seen by the observer, and $\theta$ corresponds to the average shift of the entire image from the true direction to the source. Below we numerically compute these angles using Eq.~(\ref{eq:angles}) for the different configurations of the magnetic field.

\subsection{Approximation of uncorrelated trajectories}
\label{sec:approximation_uncorr}

We begin with the simplest case of uncorrelated trajectories. In this approximation, each UHECR trajectory is treated as completely independent of the others. Technically, we implement this by recalculating the IGMF with a new seed for each particle while keeping the source and observer positions fixed. For particles emitted by a point source, this approximation is not entirely realistic, as particles that are initially closely aligned propagate through nearly the same magnetic field at short distances from the source. However, it becomes valid at large distances from the source, where the particles lose memory of their initial conditions.  Moreover, in this approximation, the angles $\alpha$, $\delta$, and $\theta$ can be computed analytically, which is useful for validating numerical results. 
\begin{figure*}
    \begin{minipage}[t]{0.483\textwidth}
        \includegraphics[width=1\linewidth]{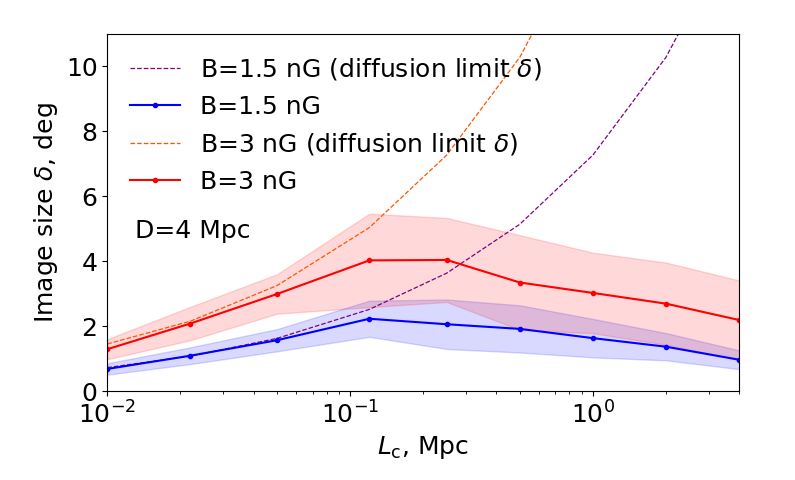}
        \caption{Angular size $\delta$ of the UHECR source image as a function of the IGMF correlation length for the realistic case where particles are emitted into the same realization of the IGMF (see Sec.~\ref{sec:image_shift} for details). Solid lines indicate the mean image size while the shaded regions indicate its standard deviation. Both the mean and standard deviation were calculated from 30 different IGMF realizations. For comparison, dashed lines show the analytical expectation for the case of uncorrelated trajectories, based on Eq.~(\ref{eq:angles_theor}). The distance to the source was set to $D=4$~Mpc.}
        \label{fig:SpotSizeL}
    \end{minipage}
    \hfill
    \begin{minipage}[t]{0.483\textwidth}
        \includegraphics[width=1\linewidth]{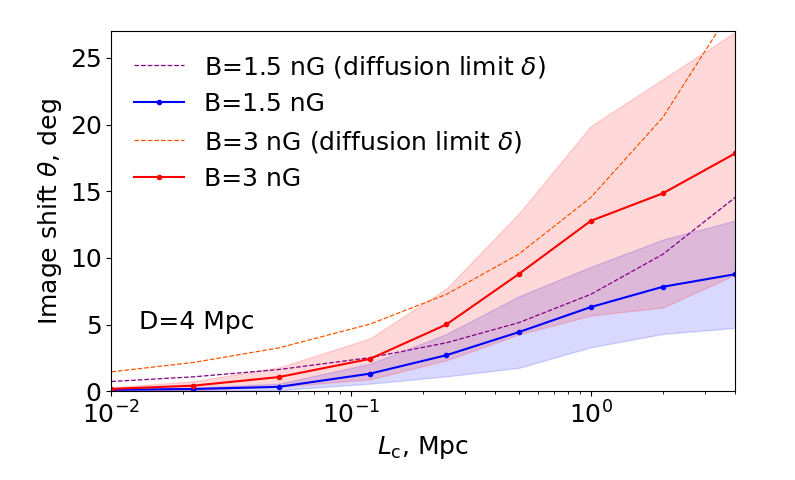}
        \caption{Same as Fig.~\ref{fig:SpotSizeL}, but here the solid lines and shaded regions represent the mean and standard deviation of the image shift $\theta$. The dashed lines are the same as in Fig.~\ref{fig:SpotSizeL} and show the analytical expectation of the image size $\delta$ for uncorrelated trajectories, based on Eq.~(\ref{eq:angles_theor}).}
        \label{fig:DirectionChangeL}
    \end{minipage}
\end{figure*}

\begin{figure*}
    \begin{minipage}[t]{0.483\textwidth}
        \includegraphics[width=1\linewidth]{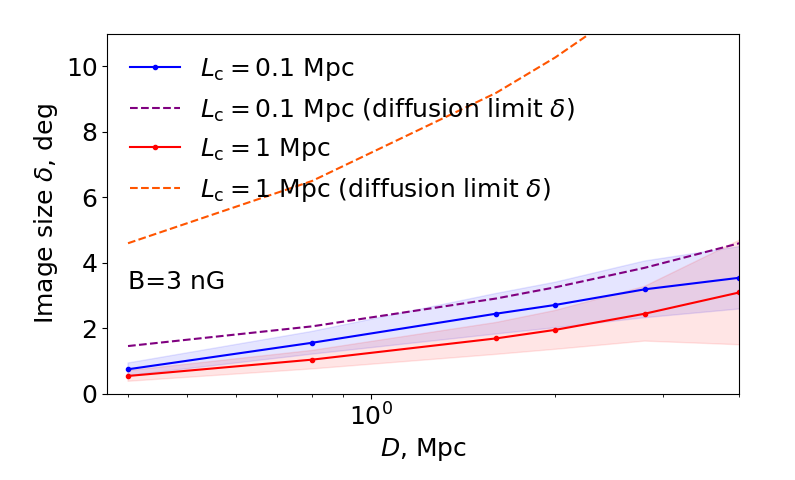}
        \caption{Angular size $\delta$ of the UHECR source image as a function of the distance to the source for the realistic case where particles are emitted into the same realization of the IGMF (see Sec.~\ref{sec:image_shift} for details). Different colors correspond to the different correlation lengths of the IGMF. The IGMF strength was fixed to $B=3$~nG. As in Fig.~\ref{fig:SpotSizeL}, the solid lines and shaded regions represent the mean and standard deviation of the image sizes, while dashed line corresponds to Eq.~(\ref{eq:angles_theor}).}
        \label{fig:SpotSizeD}
    \end{minipage}
    \hfill
    \begin{minipage}[t]{0.483\textwidth}
        \includegraphics[width=1\linewidth]{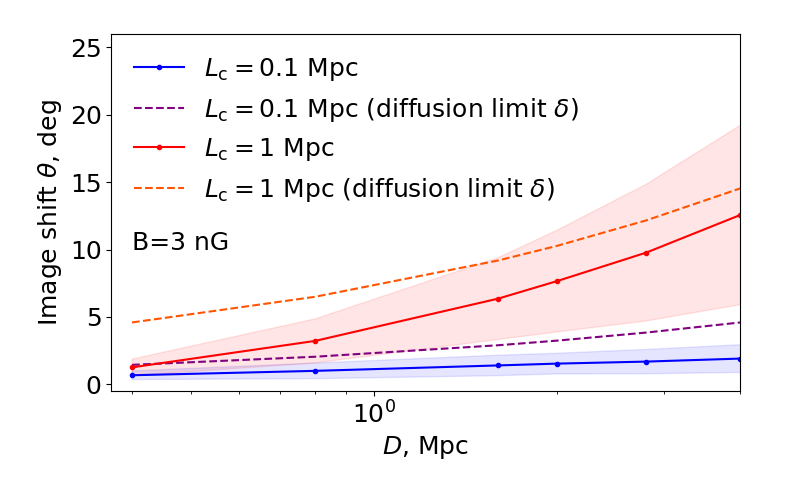}
        \caption{Same as Fig.~\ref{fig:SpotSizeD} but here solid and dashed lines indicate the mean and standard deviation of the image shift $\theta$.}
        \label{fig:DirectionChangeD}
    \end{minipage}
\end{figure*}

The classical analytical results for the case of uncorrelated trajectories were obtained in \cite{TinyakovTkachev, Achterberg1999, CapriniGabici}. Reproducing them for the magnetic field spectrum described above and assuming that the propagation distance is much greater than the maximum scale of the IGMF $D \gg L_\mathrm{max}$, we obtain
\begin{equation}\label{eq:angles_theor}
    \begin{cases}
            \alpha = 3.74^\circ\cdot Z \left[\frac{B}{1\,\mathrm{nG}}\right]\left[\frac{E}{10^{19}\,\mathrm{eV}}\right]^{-1}\left[\frac{D}{10\,\mathrm{Mpc}}\right]^\frac{1}{2}\left[\frac{L_\mathrm{max}}{500\,\mathrm{kpc}}\right]^\frac{1}{2} \\ 
            \delta = \alpha/\sqrt{3} \\
            \theta = 0.
    \end{cases}
\end{equation}
Note that while the expression for $\alpha$ depends on  assumed magnetic field spectrum, the relation between the image size $\delta$ and $\alpha$ is general and holds for any magnetic field, in the regime $D\gg L_\mathrm{max}$~\footnote{The regime $D\gg L_\mathrm{max}$ may not practically exist, for example, for a magnetic field with a scale-invariant spectrum, where, for any distance $D$, the significant energy density of the magnetic field will be contained in modes with wavelengths $L > D$.}. The absence of image shift ($\theta=0$) is also a general result and follows from the presence of axial symmetry. For the sake of readability, in the Appendix, we re-derive the expressions for $\alpha$ and $\delta$ for a magnetic field with an arbitrary spectrum, following \cite{TinyakovTkachev, Achterberg1999, CapriniGabici}.

A comparison between the theoretical predictions from Eq.~(\ref{eq:angles_theor}) and the results inferred from the numerical simulations using Eq.~(\ref{eq:angles}) are shown in Fig.~\ref{fig:AvgRandomSpotSize}. One can see that both, the particles deflection angle $\alpha$ and size of the image $\delta$, calculated numerically, closely match the theoretical curves for $L_\mathrm{c} \lesssim D/5$, when the random walk regime fully developed. Additionally, the shift of the image as a whole is zero ($\theta = 0$), as expected. The perfect agreement between the theory and the numerical simulations validates the correctness of the image construction algorithm, which allows us to use it for the more realistic case of correlated trajectories, discussed in the next Section.

\subsection{Shift of source images in the turbulent IGMF}
\label{sec:image_shift}
In this section, we repeat the simulation from the previous section, with the only difference that all particles are now propagated through the same realization of the magnetic field. As we show below, this leads to a non-zero angle $\theta$ defined in Eq.~(\ref{eq:angles}), resulting in a shift of the entire source image. On the other hand, this introduces a dependence of the results on the specific positions of the source and the observer within a given IGMF realization. To investigate this effect, we perform 30 simulations, each with its own realization of the IGMF. 

The results of the simulations for a source at a distance of $D=4$~Mpc are shown in Figs.~\ref{fig:SpotSizeL} and \ref{fig:DirectionChangeL}. Fig.~\ref{fig:SpotSizeL} shows the dependence of the image size $\delta$ on the IGMF correlation length $L_\mathrm{c}$. The size of the image approaches the theoretical prediction for uncorrelated trajectories, given by Eq.~(\ref{eq:angles_theor}), only for small correlation lengths, $L_\mathrm{c} \lesssim 0.1$~Mpc. This regime corresponds to the propagation of particles through approximately $D / L_\mathrm{c} \gtrsim 100$ magnetic domains.  In contrast, for larger correlation lengths ($L_\mathrm{c} > 0.1$~Mpc), the image size remains small and even decreases with increasing $L_\mathrm{c}$. One can compare this with the case of uncorrelated trajectories (see Fig.~\ref{fig:AvgRandomSpotSize}),where the image size begins to follow the theoretical prediction after passing through only $\sim 5$ magnetic domains.

On the other hand, as seen in {Fig.~\ref{fig:DirectionChangeL}}, there is a non-zero shift of the entire source image. At small correlation lengths, the shift tends to zero, but increases rapidly as $L_\mathrm{c}$ grows. Notably, when the correlation length is in the range $\sim 0.1–1$~Mpc, the image shift, averaged over 30 realizations, approaches the expected value for the image size in the case of uncorrelated trajectories. A similar behavior is observed in Figs.~\ref{fig:SpotSizeD} and \ref{fig:DirectionChangeD}, which show the image size $\delta$ and the image shift $\theta$ as functions of the distance to the source. The image size is smaller than in the case of uncorrelated trajectories, while the image shift is significant.

This dependence of the image size and shift on the correlation length and distance to the source can be explained by the fact that the particles propagate through the same realization of the magnetic field. Close to the source, particles emitted in similar initial directions travel through almost the same magnetic field, causing the same deflection for all particles. This leads to a shift of the entire image while its size remains small. At the same time, the average shift angle follows the expected value of the image size. Only at a large distance, when particles pass through the $\sim 100$ magnetic domains and forget their initial conditions, the approximation of uncorrelated trajectories is restored.

\subsection{Application to the Centaurus region excess}
\label{sec:cen_a}
As a specific application of UHECR source image shifts induced by the IGMF, we consider the interpretation of the UHECR hotspot in the Centaurus region, observed at energies $E > 4 \cdot 10^{19}$~eV \cite{PierreAuger:2022axr,PierreAuger:2023fcr}. This is the most significant intermediate-scale anisotropy detected by the Auger. The Centaurus region, as defined in the Auger analysis, contains sources belonging to two potentially important classes of UHECR accelerators. First, it includes the nearest radio galaxy, Centaurus A. Second, it hosts two starburst galaxies: NGC~4945 and M83. Without accounting for deflections in the GMF the last two sources (NGC~4945 and M83) make the largest contribution to the correlations of the arrival directions of Auger events with starburst galaxies~\cite{PierreAuger:2018qvk}. On the other hand, the results of numerical simulations indicate that Cen~A may be the source of a considerable fraction of the observed UHECR flux~\cite{Eichmann:2017iyr,Matthews:2018laz,Mollerach:2021ifa}. 

As shown in \cite{Farrar:2012gm, Keivani:2014kua, Allard:2023uuk}, accounting for deflections in the GMF can significantly complicate the picture by shifting the apparent positions of potential sources. For instance, in the \texttt{JF12} GMF model~\cite{JF_GMF_1, JF_GMF_2}, the image of Centaurus A is displaced by approximately $20^\circ$ from the location of the observed excess. In the following, we revisit the problem of the origin of the Cen~A excess, now also considering a possible additional deflection caused by the IGMF. We adopt the \texttt{JF12} model as our reference GMF; however, our conclusions are not sensitive to the specific choice of the GMF model. Similar results can be obtained using newer \texttt{UF23} and \texttt{KST24} models. Motivated by the UHECR mass composition measurements by Auger{~\cite{PierreAuger:2023gmj}}, which suggest that the highest energy flux is dominated by a narrow range of rigidities, we consider monochromatic particles with a rigidity of $10$~EV. {However, we note that the interpretation of dipole anisotropy observed by Auger involves lower values of rigidity~\cite{Rossoni:2025vsk, Bister:2023icg}. We have verified that our results are robust to slight changes in rigidity. Repeating the simulations for particles with $R=5$~EV and $R=7$~EV, we confirmed that the resulting images are similar to those obtained for $R=10$~EV.}

\begin{figure*}
    \includegraphics[width=0.49\linewidth]{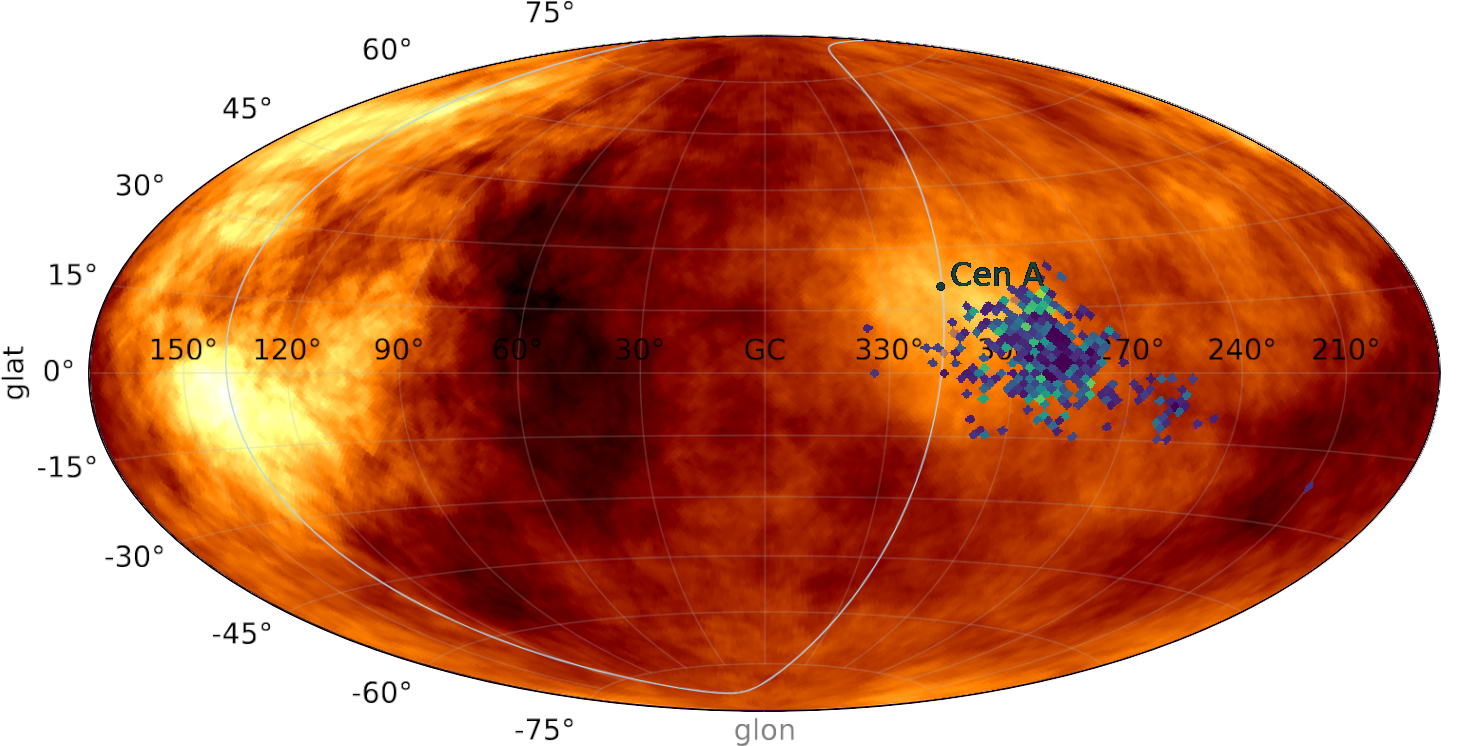}
    \hfill
    \includegraphics[width=0.49\linewidth]{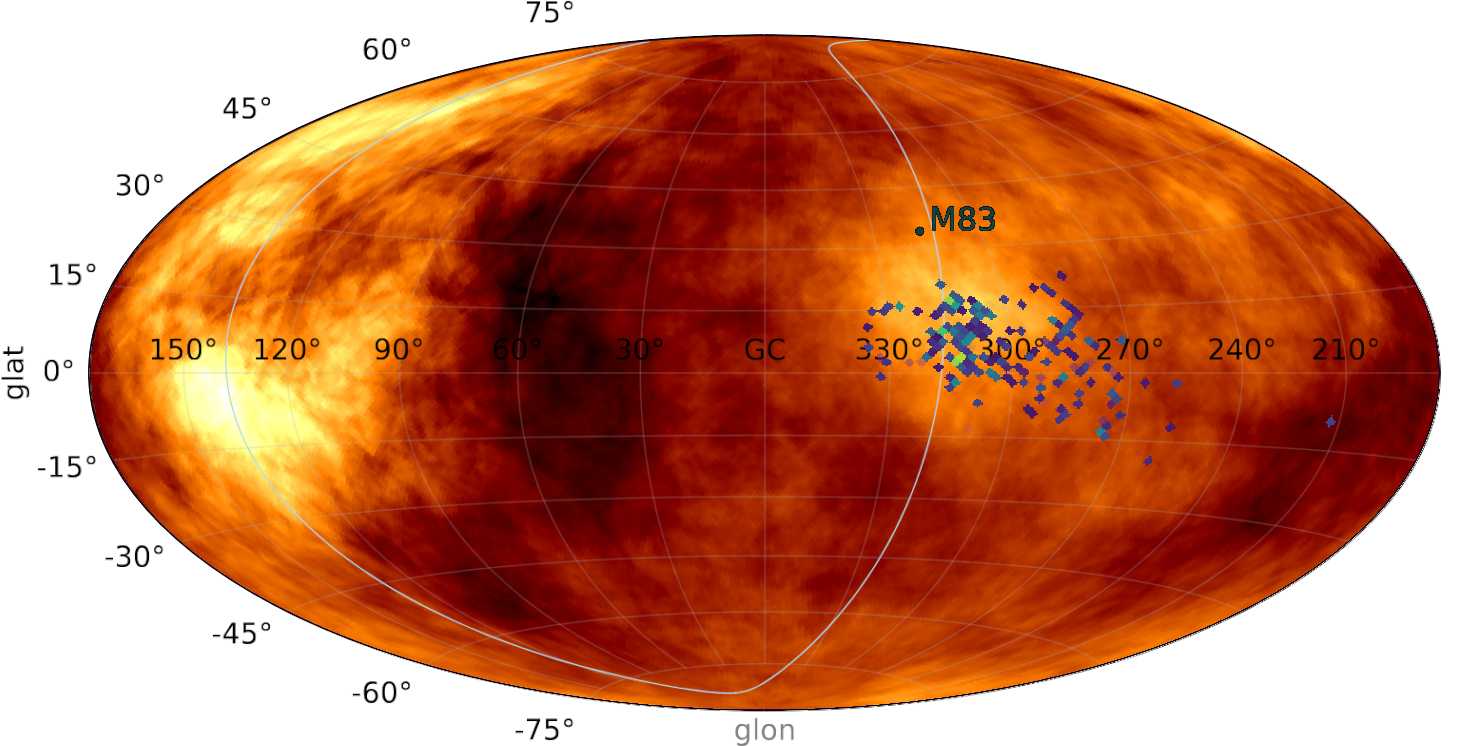}
    \caption{Model prediction of the arrival directions of the carbon nuclei with $E = 60$ EeV  after propagation in the  IGMF with strength 1.5 nG and coherence length  $L_\mathrm{c}=0.05$ Mpc and after propagation in the  GMF in JF12 model. The arrival directions of the events are superimposed on the UHECR full-sky flux map measured by Auger and TA~\cite{PierreAuger:2023mvf} with the hotspots indicated by bright regions { on hot colormap}. { The density of simulated events is shown as green-blue colormap which is plotted above the hot colormap.} For the left figure Cen~A is the source of the observed UHECR.  {For the} right figure M83 is the source of the observed UHECR. {  TA hotspots are brighter than the Auger Cen A one, while the  the latter has higher statistical significance than the former due to larger exposure of Auger.}}
    \label{fig:CenAM83_onlyGMF}
\end{figure*}
\begin{figure*}
    \includegraphics[width=0.49\linewidth]{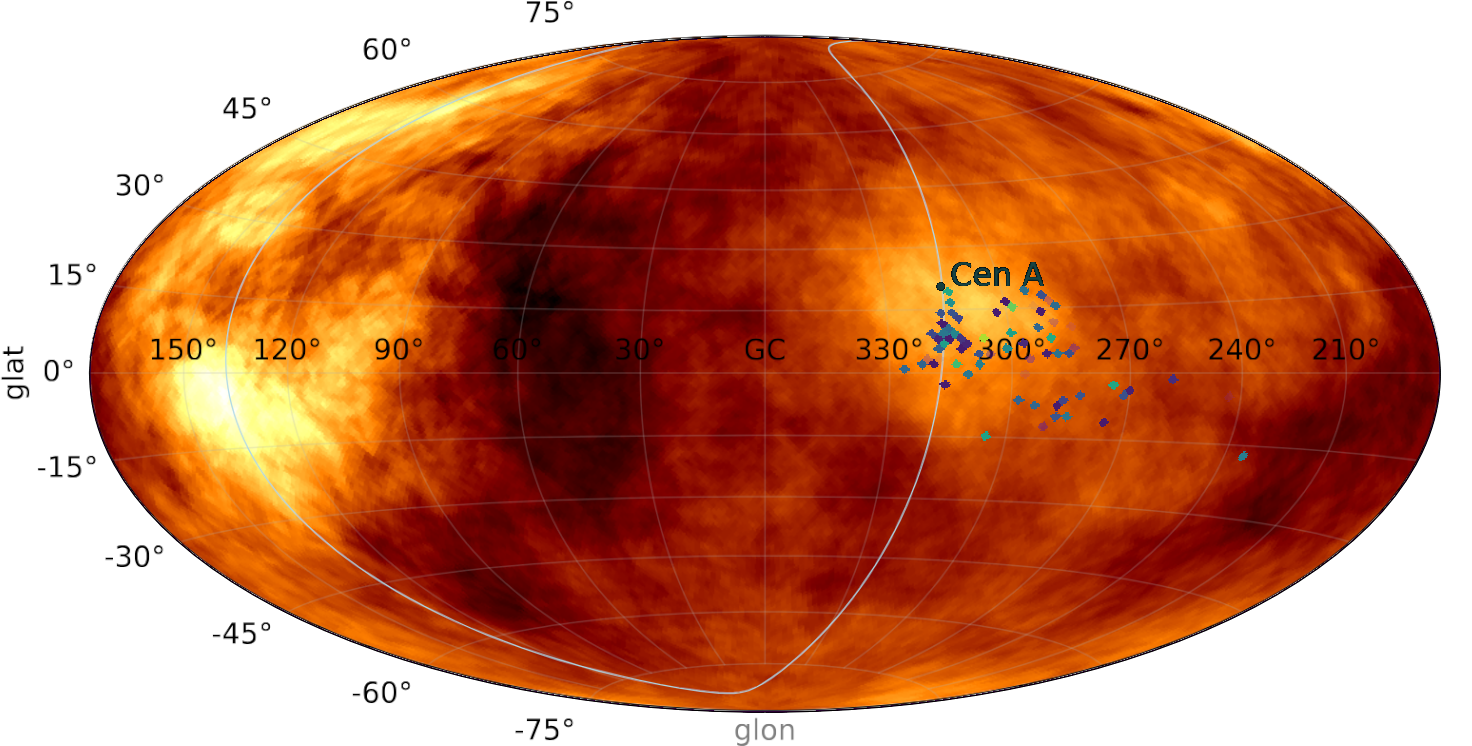}
    \hfill
    \includegraphics[width=0.49\linewidth]{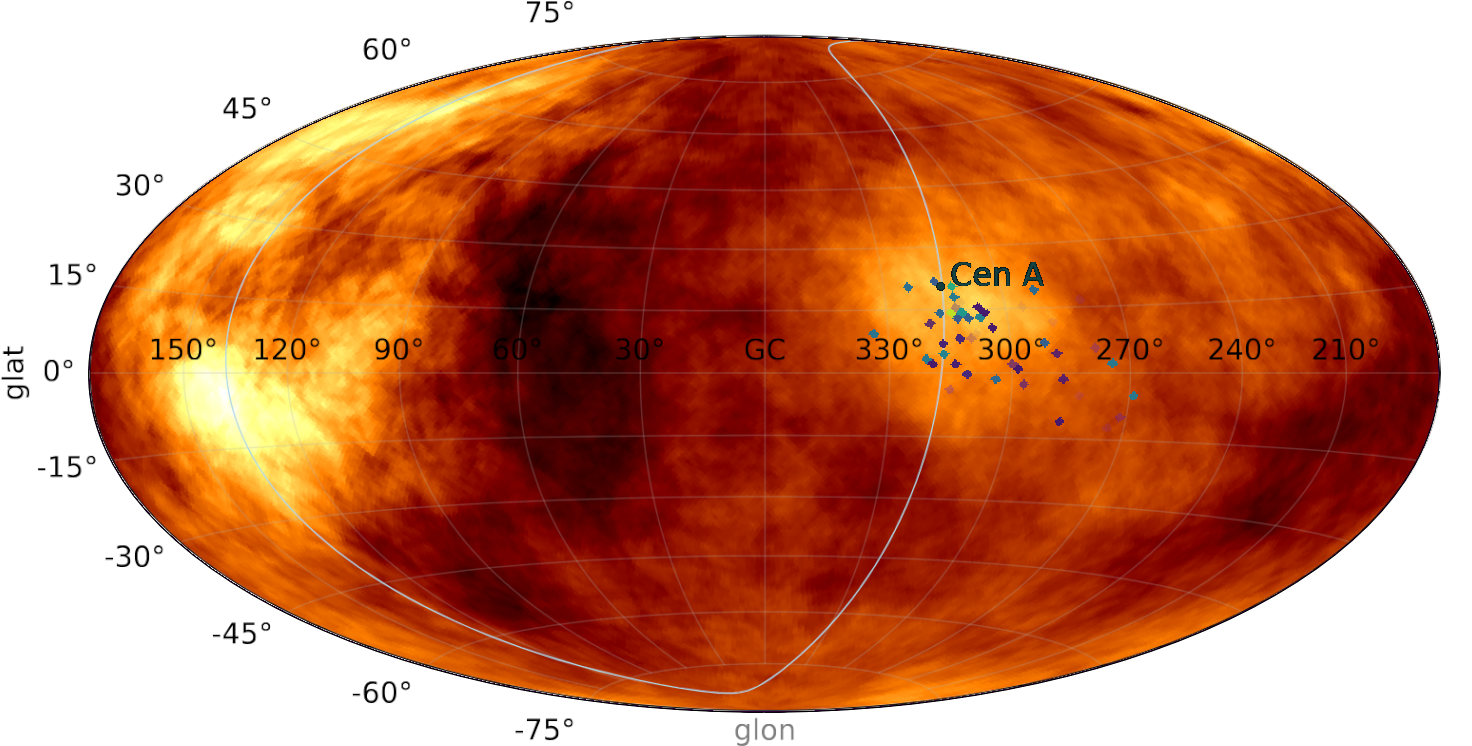}
    \caption{Arrival directions of $E=60$~EeV carbon nuclei from Cen~A after propagation through the GMF and IGMF. The left panel corresponds to an IGMF with $B=1.5$~nG and $L_\mathrm{c}=3.35$~Mpc, while the right panel shows the case with $B=3$~nG and $L_\mathrm{c}=0.3$~Mpc. {In both cases, a particular realization of the IGMF was chosen in which the image shift induced by the GMF is compensated by the shift in the IGMF, so that the resulting image remains centered on the observed hotspot.}}
    \label{fig:CenA_GMFandIGMF}
\end{figure*}

First, we construct the images of M83 and Cen~A as they would appear on Earth, accounting only for deflections in the GMF. To do this, we use the Galactic magnetic lensing method implemented in CRPropa3~\cite{AlvesBatista:2016vpy, AlvesBatista:2022vem}, which enables efficient forward tracking of UHECRs by making use of backtracked precomputed magnetic lenses for the \texttt{JF12} GMF model. These lenses incorporate the coherent, striated, and turbulent components of the \texttt{JF12} model~\cite{JF_GMF_1, JF_GMF_2}.

The image of Cen~A, as observed at Earth, is shown in the left panel of Fig.~\ref{fig:CenAM83_onlyGMF}. In the \texttt{JF12} model, the coherent GMF component shifts particles with rigidity of $10$~EV (corresponding to carbon nuclei at energies $E \gtrsim 4 \cdot 10^{19}$~eV) by approximately $20^\circ$ from the position of the excess toward the Galactic plane and lower Galactic longitudes. The turbulent component causes an additional angular spread around the image center. In contrast, the image of M83 perfectly fits the position of the excess region as shown on the right panel of Fig.~\ref{fig:CenAM83_onlyGMF}.

It is important to note that the GMF structure in the direction of the Cen~A excess remains poorly constrained, and there are significant differences between different GMF models, as discussed in~\cite{2025arXiv250116158K}. Thus, given the current uncertainties in GMF modeling, no definitive conclusion can be drawn regarding M83 as the source of the observed excess. In addition, the IGMF with relatively large correlation length $L_\mathrm{c} \gtrsim 0.1$~Mpc introduces an additional uncertainty. This is illustrated in Fig.~\ref{fig:CenA_GMFandIGMF}, which shows the images of Centaurus A after accounting for both GMF and IGMF deflections. As seen in the figure, the resulting images of Cen~A overlap with the position of the excess observed. 

This alignment is achieved by selecting a particular IGMF realization in which the image of Cen~A is first deflected toward the position of M83 and then further shifted by the GMF to match the location of the observed excess. Thus, the combined effect of IGMF and GMF still allows the Cen~A to be the source of excess. 

{Since the proposed idea relies on the mutual cancellation of the shifts induced by the GMF and IGMF, it is instructive to estimate the probability of this occurring by chance for a given GMF model. The angular distance between M83 and the center of Cen A is approximately $12^\circ$. According to our simulations, the probability of an image shift exceeding $10^\circ$ is about 13\% for an IGMF with $B=3$~nG and $L_\mathrm{c}=0.3$~Mpc. An additional degree of freedom is provided by rotating the IGMF around a straight line connecting the observer and the source, and the probability that the shift occurs in the required direction is approximately $15^\circ/360^\circ \approx 4\%$, where $15^\circ$ is approximately the half-size of Cen A excess. Thus, the total probability for the case shown in the right panel of Fig.~\ref{fig:CenA_GMFandIGMF} to occur by chance is roughly 0.5\%.}

\section{Conclusions}
\label{sec:conclusions}

We studied the propagation of UHECRs in the focusing regime within a turbulent intergalactic magnetic field, from their sources to the Milky Way Galaxy. In this regime, the UHECRs are initially deflected by similar magnetic-field structures as they pass through the first few coherence lengths of the turbulent IGMF. As a result, within a distance of approximately ten coherence lengths from the source, UHECRs tend to be focused into circular overdense regions across most of the sky. These overdensities follow the local structure of $\mathrm{rot}\vec{B}$  in the initial cells of the turbulent field \cite{dolgikh2022causticlike, DOLGIKH20245295}. Consequently, the standard assumption of simplified diffusion theory — that each cosmic ray propagates through an independent realization of the IGMF — is no longer valid. 

The most important result is that in the focusing mode, at a distance from the source less than one hundred magnetic field coherence lengths ($D < 100,L_\mathrm{c}$), the standard expectation of diffusion theory, according to which cosmic rays are isotropically distributed around their sources, is violated. Contrary to the predictions of diffusion theory, the observed images of UHECR sources are both smaller than expected and can be significantly displaced from the actual source positions, see Figs.~\ref{fig:SpotSizeL} and \ref{fig:DirectionChangeL}.

In this paper, we used this focusing mode to search for a possible source of excess cosmic rays - a hotspot near Cen~A. We took into account both the IGMF and the GMF contribution to deflections of UHECR. We assumed that IGMF is turbulent with a strength of $B \sim 1$~nG and coherence length $L_\mathrm{c} < 5$~Mpc. We combined the propagation of UHECR in IGMF with the propagation in the Milky Way Galaxy, for which we used the \texttt{JF12} model ~\cite{JF_GMF_1,JF_GMF_2}.

We obtained three important results. First, {in agreement with the results of \cite{Farrar:2012gm,Keivani:2014kua,Farrar:2017lhm} in the JF12 model}, cosmic ray carbon nuclei with energies of 
$E=60$~EeV originating from Cen~A are deflected toward the Galactic plane, resulting in an arrival direction distribution at Earth that is inconsistent with the observed excess near Cen~A. Second, if instead of Cen~A we consider M83 as the source of the same UHECRs, propagation through the GMF deflects the particles toward the observed position of Cen~A excess. Third, when the IGMF has a coherence length comparable to or smaller by an order of magnitude than the source distance, a focusing regime arises. In this regime, Cen~A can still be the true source of the UHECRs observed in its vicinity. Specifically, UHECRs are first deflected by the IGMF towards a position near M83, and then redirected by the GMF back toward Cen~A. This scenario highlights the critical roles of both the GMF and IGMF in interpreting the observed arrival directions of UHECRs and in identifying their sources.

\section*{Acknowledgments}
We are grateful to Peter Tinyakov for useful discussions. The numerical modeling of the propagation of ultra-high-energy cosmic rays under the influence of large-scale cosmic magnetic fields (K.D.) is supported by the Russian Science Foundation grant 25-12-00111. The work of K.D. is supported by the Basis Foundation fellowship (grant No. 22-1-5-93-1). The work of D.S. is supported by the French National Research Agency (ANR) grant ANR-24-CE31-4686. The work of A.K. is supported by the IISN project No. 4.4501.18. Some of the results in this paper have been derived using the healpy and HEALPix packages.

\appendix*
\section{Particle deflections and source image size in turbulent magnetic field}
In this Appendix, we derive analytical expressions for the rms deflection angle of particles, $\alpha$, and the angular size of the source image, $\delta$, after propagation in a turbulent magnetic field. The equations are obtained assuming that the deflection angle is small $\alpha \ll 1$ and that the particle trajectories are statistically independent (i.e., the approximation of uncorrelated trajectories). Our derivation closely follows the approach of \cite{TinyakovTkachev} and \cite{CapriniGabici}, although similar results were previously obtained in \cite{Achterberg1999} using a different method.

A homogeneous turbulent magnetic field $B$ can be characterized by its correlation function, defined as $\zeta_{ij}(\mathbf{x}) = \langle B_i(\mathbf{x_0})B_j(\mathbf{x_0 +x})\rangle$. To express this correlation function, we introduce its representation in Fourier space and find its relation to the real-space correlator. Assume that
\begin{equation*}
    \begin{cases}
        B_i(\mathbf{x}) = \frac{1}{(2\pi)^3}\int\mathrm{d}^3k\,B_i(\mathbf{k}) e^{i\mathbf{k}\mathbf{x}}\vspace{0.1cm}\\ 
        \langle B_i(\mathbf{k}) B^{*}_j(\mathbf{q}) \rangle = (2\pi)^3\,\frac{1}{2} (\delta_{ij} - \hat{k}_i \hat{k}_j) \, P_B(k) \,\delta(\mathbf{k} - \mathbf{q}) \end{cases}
\end{equation*}
where $(\delta_{ij} - \hat{k}_i \hat{k}_j)$ is required to follow the condition $\mathrm{div}\mathbf{B} = 0$, $\hat{\mathbf{k}} = \mathbf{k}/k$, and $P_B(k)$ is the power spectrum of the magnetic field. Substituting this into the definition of the correlation function, we obtain
\begin{equation}\label{eq:corr}
    \zeta_{ij}(\mathbf{x}) = \frac{1}{(2\pi)^3} \frac{1}{2} \int\mathrm{d}^3k\,(\delta_{ij} - \hat{k}_i \hat{k}_j)\, P_B(k)\,e^{-i\mathbf{kx}}.
\end{equation}
The normalization of $P_\mathrm{B}(k)$ is determined by requiring that the rms magnetic field strength equals a given value $B_0$
\begin{equation}
    B_0^2 = \langle B_i(\mathbf{x}) B_i(\mathbf{x}) \rangle = \zeta_{ii}(0).
\end{equation}
Using $\delta_{ii} = 3$ and $\hat{k}_i \hat{k}_i = 1$ we arrive at the normalization condition:
\begin{equation}\label{eq:norm}
    B_0^2 = \frac{1}{(2\pi)^3}\int \mathrm{d}^3k\,P_B(k).
\end{equation}

Given these definitions, the rms deflection angle of a particle $\alpha$ can be derived starting from the Lorentz equation of motion:
\begin{equation}
    \frac{\mathrm{d}\mathbf{p}}{\mathrm{d}t} = \frac{e}{c}\cdot[\mathbf{v}\times\mathbf{B}].
\end{equation}
Assuming that the deflection is small, we approximate the velocity as $\mathbf{v} \approx c\cdot\mathbf{\hat{n}}$.  Further, assuming that the particle propagates along the $z$-axis, so that $\mathbf{\hat{n}} = (0, 0, 1)$, and considering motion from $z = 0$ to $z = D$, we obtain
\begin{equation}
    \langle\alpha^2\rangle = \left(\frac{e}{E}\right)^2 \int\limits_0^D \mathrm{d}z_1 \int\limits_0^D \mathrm{d}z_2 \, [\delta_{kq} - \hat{n}_k \hat{n}_q]\, \langle B_k (z_1) B_q (z_2) \rangle
\end{equation}
where $\langle\cdot\rangle$ denote averaging over different realizations of the magnetic field. 

Substituting Eq.~(\ref{eq:corr}) into these equations and performing the analytical integrations, we obtain the most general expression for the deflection angle \footnote{Note that in \cite{CapriniGabici}, there is a typo in the signs of two out of three terms; however, this does not affect the final results.}:
\begin{multline}\label{eq:general}
    \langle\alpha^2\rangle = D\left(\frac{e}{E}\right)^2\,\frac{2}{(2\pi)^2}\,\int\limits_{0}^{\infty}\mathrm{d}k\, \,kP_B(k)\times \\ \left(\frac{D^2k^2\,\mathrm{Si}(Dk) + Dk\cos{(Dk)} - \sin{(Dk)}}{D^2k^2}\right).
\end{multline}
The Eq.~(\ref{eq:general}) for the rms deflection angle $\alpha$ is valid for any distance $D$ and any magnetic field power spectrum. The only assumption made so far is that $\alpha$ is small. In \cite{CapriniGabici}, this integral was evaluated analytically for a broken power-law spectrum in two limiting cases $Dk_0 \ll 1$ and $Dk_0 \gg 1$ where $k_0$ corresponds to the position of the break in the spectrum. For more complex spectra, it can be computed numerically.

For the magnetic field spectrum given by Eq.~\ref{eq:igmf_spec} and used in this paper, the integral in Eq.~(\ref{eq:general}) can also be evaluated analytically in the limit where the propagation distance is much larger than the largest scale of the field, i.e. $Dk_\mathrm{min} \gg 1$ or equivalently $D/L_\mathrm{max} \gg 1$:
\begin{equation}
    \langle\alpha^2\rangle = \frac{D L_\mathrm{max}}{R^2_\mathrm{L}}\,\left[\frac{1}{4}\frac{(m-3)}{(m-2)}\,\,\frac{(1 - R_k^{m-2})}{(1 - R_k^{m-3})}\right].
\end{equation}
Here $R_k = L_\mathrm{min}/L_\mathrm{max}$ is the ratio of the minimum to maximum scales of the spectrum, and $R_\mathrm{L} = E/eB$ is the Larmor radius of the particle. Assuming a sufficiently broad spectrum with $R_k \ll 1$ ($R_k = 0.01$ for the spectrum in Eq.~(\ref{eq:igmf_spec}) and substituting the Kolmogorov slope $m=11/3$, one obtains
\begin{equation}
    \langle\alpha^2\rangle = \frac{D L_\mathrm{max}}{10R^2_\mathrm{L}}\,\left[1 + R_k^{2/3}\right]
\end{equation}
to the leading order in $R_k$. For the values of $B$, $E$, $D$, and $L_\mathrm{max}$ relevant to UHECR propagation, with expression reduces to the one from Eq.~\ref{eq:angles_theor}.

To compute the rms angular size of the source, $\delta$ one starts with Eq.~(\ref{eq:im_size}). In the approximation of uncorrelated trajectories, the definition of $\delta$ in Eq.~(\ref{eq:im_size}) is equivalent to:
\begin{equation}
    \cos\delta = \left(\mathbf{n} \cdot \frac{\mathbf{r}}{r}\right)
\end{equation}
where $\mathbf{r}$ is the particle's radius-vector. Averaging over different realizations of the magnetic field and assuming as always that $\delta \ll 1$ one obtains to leading order:
\begin{equation}\label{eq:size_init}
    \langle\delta^2\rangle = \langle n_x^2\rangle + \langle n_y^2\rangle - 2\left[\frac{\langle r_xn_x\rangle}{D} + \frac{\langle r_yn_y\rangle}{D}\right] + \frac{\langle r_x^2\rangle}{D^2} + \frac{\langle r_y^2\rangle}{D^2}.
\end{equation}
Note, that
\begin{equation}
    \langle n_x^2\rangle + \langle n_y^2\rangle = \langle\alpha^2\rangle.
\end{equation}
Similarly, 
\begin{multline}
    \langle r_x n_x\rangle + \langle r_y n_y\rangle = \\ = \int\limits_0^D\mathrm{d}z_1\,\left[\left(\frac{e}{E}\right)^2 \int\limits_0^{z_1} \mathrm{d}z_2 \int\limits_0^D \mathrm{d}z_3\, [\epsilon_{ijk} \hat{n}_j B_k (z_2) \,\epsilon_{ipq} \hat{n}_p B_q (z_3)]\right]
\end{multline}
So, that
\begin{equation}
    \langle r_x n_x\rangle + \langle r_y n_y\rangle = \frac{D}{2}\langle \alpha^2\rangle
\end{equation}
Finally,
\begin{equation}\label{eq:r_avg}
    \langle r_x^2\rangle + \langle r_y^2\rangle = \frac{D^2}{3}\langle \alpha^2\rangle
\end{equation}
So, 
\begin{equation}
    \langle \delta^2\rangle = \frac{\langle\alpha^2\rangle}{3}
\end{equation}

\bibliographystyle{apsrev4-1}  
\bibliography{refs.bib}

@article{Harari:2002dy,
    author = "Harari, Diego and Mollerach, Silvia and Roulet, Esteban and Sanchez, Federico",
    title = "{Lensing of ultrahigh-energy cosmic rays in turbulent magnetic fields}",
    eprint = "astro-ph/0202362",
    archivePrefix = "arXiv",
    doi = "10.1088/1126-6708/2002/03/045",
    journal = "JHEP",
    volume = "03",
    pages = "045",
    year = "2002"
}

@article{PierreAuger:2015eyc,
    author = "Aab, Alexander and others",
    collaboration = "Pierre Auger",
    title = "{The Pierre Auger Cosmic Ray Observatory}",
    eprint = "1502.01323",
    archivePrefix = "arXiv",
    primaryClass = "astro-ph.IM",
    reportNumber = "FERMILAB-PUB-15-034-AD-AE-CD-TD",
    doi = "10.1016/j.nima.2015.06.058",
    journal = "Nucl. Instrum. Meth. A",
    volume = "798",
    pages = "172--213",
    year = "2015"
}

@ARTICLE{2012NIMPA.676...54T,
       author = {{Tokuno}, H. and {Tameda}, Y. and {Takeda}, M.
                  and others},
        title = "{New air fluorescence detectors employed in the Telescope Array experiment}",
      journal = {Nuclear Instruments and Methods in Physics Research A},
         year = 2012,
        month = jun,
       volume = {676},
        pages = {54-65},
          doi = {10.1016/j.nima.2012.02.044},
archivePrefix = {arXiv},
       eprint = {1201.0002},
 primaryClass = {astro-ph.IM},
       adsurl = {https://ui.adsabs.harvard.edu/abs/2012NIMPA.676...54T}
}

@ARTICLE{2012NIMPA.689...87A,
       author = {{Abu-Zayyad}, T. and {Aida}, R. and {Allen}, M. and others},
        title = "{The surface detector array of the Telescope Array experiment}",
      journal = {Nuclear Instruments and Methods in Physics Research A},
         year = 2012,
        month = oct,
       volume = {689},
        pages = {87-97},
          doi = {10.1016/j.nima.2012.05.079},
archivePrefix = {arXiv},
       eprint = {1201.4964},
 primaryClass = {astro-ph.IM},
       adsurl = {https://ui.adsabs.harvard.edu/abs/2012NIMPA.689...87A}
}

@ARTICLE{2017Sci...357.1266P,
       author = {{Pierre Auger Collaboration} and {Aab}, A. and {Abreu}, P. and {Aglietta}, M. and others},
        title = "{Observation of a large-scale anisotropy in the arrival directions of cosmic rays above 8 {\texttimes} {}10$^{18}$ eV}",
      journal = {Science},
         year = 2017,
        month = sep,
       volume = {357},
       number = {6357},
        pages = {1266-1270},
          doi = {10.1126/science.aan4338},
archivePrefix = {arXiv},
       eprint = {1709.07321},
 primaryClass = {astro-ph.HE},
       adsurl = {https://ui.adsabs.harvard.edu/abs/2017Sci...357.1266P}
}

@ARTICLE{2024ApJ...976...48A,
       author = {{Abdul Halim}, A. and {Abreu}, P. and {Aglietta}, M. and others},
        title = "{Large-scale Cosmic-ray Anisotropies with 19 yr of Data from the Pierre Auger Observatory}",
      journal = {\apj},
         year = 2024,
        month = nov,
       volume = {976},
       number = {1},
          eid = {48},
        pages = {48},
          doi = {10.3847/1538-4357/ad843b},
archivePrefix = {arXiv},
       eprint = {2408.05292},
 primaryClass = {astro-ph.HE},
       adsurl = {https://ui.adsabs.harvard.edu/abs/2024ApJ...976...48A}
}

@INPROCEEDINGS{2022icrc.confE.375T,
       author = {{Tinyakov}, P. and {Kim}, J. and {Kuznetsov}, M. and others},
        title = "{The UHECR dipole and quadrupole in the latest data from the original Auger and TA surface detectors}",
    booktitle = {37th International Cosmic Ray Conference},
         year = 2022,
        month = mar,
          eid = {375},
        pages = {375},
          doi = {10.22323/1.395.0375},
archivePrefix = {arXiv},
       eprint = {2111.14593},
 primaryClass = {astro-ph.HE},
       adsurl = {https://ui.adsabs.harvard.edu/abs/2022icrc.confE.375T}
}

@ARTICLE{TinyakovTkachev,
       author = {{Tinyakov}, P.~G. and {Tkachev}, I.~I.},
        title = "{Deflections of cosmic rays in a random component of the Galactic magnetic field}",
      journal = {Astroparticle Physics},
         year = 2005,
        month = sep,
       volume = {24},
       number = {1-2},
        pages = {32-43},
          doi = {10.1016/j.astropartphys.2005.05.003},
archivePrefix = {arXiv},
       eprint = {astro-ph/0411669},
 primaryClass = {astro-ph},
       adsurl = {https://ui.adsabs.harvard.edu/abs/2005APh....24...32T}
}

@ARTICLE{Achterberg1999,
       author = {{Achterberg}, Abraham and {Gallant}, Yves A. and {Norman}, Colin A. and {Melrose}, Donald B.},
        title = "{Intergalactic Propagation of UHE Cosmic Rays}",
      journal = {arXiv e-prints},
         year = 1999,
        month = jul,
          eid = {astro-ph/9907060},
        pages = {astro-ph/9907060},
          doi = {10.48550/arXiv.astro-ph/9907060},
archivePrefix = {arXiv},
       eprint = {astro-ph/9907060},
 primaryClass = {astro-ph},
       adsurl = {https://ui.adsabs.harvard.edu/abs/1999astro.ph..7060A}
}

@ARTICLE{CapriniGabici,
       author = {{Caprini}, C. and {Gabici}, S.},
        title = "{Gamma-ray observations of blazars and the intergalactic magnetic field spectrum}",
      journal = {\prd},
         year = 2015,
        month = jun,
       volume = {91},
       number = {12},
          eid = {123514},
        pages = {123514},
          doi = {10.1103/PhysRevD.91.123514},
archivePrefix = {arXiv},
       eprint = {1504.00383},
 primaryClass = {astro-ph.CO},
       adsurl = {https://ui.adsabs.harvard.edu/abs/2015PhRvD..91l3514C}
}

@article{PierreAuger:2023mvf,
    author = "Abdul Halim, Adila and others",
    collaboration = "Pierre Auger",
    title = "{Update on the searches for anisotropies in UHECR arrival directions with the Pierre Auger Observatory and the Telescope Array}",
    doi = "10.22323/1.444.0521",
    journal = "PoS",
    volume = "ICRC2023",
    pages = "521",
    year = "2023"
}

@ARTICLE{2024ApJ...966...71B,
       author = {{Bister}, Teresa and {Farrar}, Glennys R.},
        title = "{Constraints on UHECR Sources and Extragalactic Magnetic Fields from Directional Anisotropies}",
      journal = "Astrophys. J.",
         year = 2024,
        month = may,
       volume = {966},
       number = {1},
          eid = {71},
        pages = {71},
          doi = {10.3847/1538-4357/ad2f3f},
archivePrefix = {arXiv},
       eprint = {2312.02645},
 primaryClass = {astro-ph.HE},
       adsurl = {https://ui.adsabs.harvard.edu/abs/2024ApJ...966...71B}
}

@article{TelescopeArray:2014tsd,
    author = "Abbasi, R. U. and others",
    collaboration = "Telescope Array",
    title = "{Indications of Intermediate-Scale Anisotropy of Cosmic Rays with Energy Greater Than 57 EeV in the Northern Sky Measured with the Surface Detector of the Telescope Array Experiment}",
    eprint = "1404.5890",
    archivePrefix = "arXiv",
    primaryClass = "astro-ph.HE",
    doi = "10.1088/2041-8205/790/2/L21",
    journal = "Astrophys. J. Lett.",
    volume = "790",
    pages = "L21",
    year = "2014"
}

@ARTICLE{2021arXiv211014827T,
       author = {{Telescope Array Collaboration} and {Abbasi}, R.~U. and {Abu-Zayyad}, T. and {Allen}, M. and others},
        title = "{Indications of a Cosmic Ray Source in the Perseus-Pisces Supercluster}",
      journal = {arXiv e-prints},
         year = 2021,
        month = oct,
          eid = {arXiv:2110.14827},
        pages = {arXiv:2110.14827},
          doi = {10.48550/arXiv.2110.14827},
archivePrefix = {arXiv},
       eprint = {2110.14827},
 primaryClass = {astro-ph.HE},
       adsurl = {https://ui.adsabs.harvard.edu/abs/2021arXiv211014827T}
}

@article{PierreAuger:2018qvk,
    author = "Aab, Alexander and others",
    collaboration = "Pierre Auger",
    title = "{An Indication of anisotropy in arrival directions of ultra-high-energy cosmic rays through comparison to the flux pattern of extragalactic gamma-ray sources}",
    eprint = "1801.06160",
    archivePrefix = "arXiv",
    primaryClass = "astro-ph.HE",
    reportNumber = "FERMILAB-PUB-17-358-ND-PPD-TD",
    doi = "10.3847/2041-8213/aaa66d",
    journal = "Astrophys. J. Lett.",
    volume = "853",
    number = "2",
    pages = "L29",
    year = "2018"
}

@INPROCEEDINGS{2023EPJWC.28303002D,
       author = {{di Matteo}, A. and {Anchordoqui}, L. and {Bister}, T. and others},
        title = "{2022 report from the Auger-TA working group on UHECR arrival directions}",
    booktitle = {European Physical Journal Web of Conferences},
         year = 2023,
       series = {European Physical Journal Web of Conferences},
       volume = {283},
        month = oct,
    publisher = {EDP},
          eid = {03002},
        pages = {03002},
          doi = {10.1051/epjconf/202328303002},
archivePrefix = {arXiv},
       eprint = {2302.04502},
 primaryClass = {astro-ph.HE},
       adsurl = {https://ui.adsabs.harvard.edu/abs/2023EPJWC.28303002D}
}

@ARTICLE{2024ApJ...970...95U,
       author = {{Unger}, Michael and {Farrar}, Glennys R.},
        title = "{The Coherent Magnetic Field of the Milky Way}",
      journal = "Astrophys. J.",
         year = 2024,
        month = jul,
       volume = {970},
       number = {1},
          eid = {95},
        pages = {95},
          doi = {10.3847/1538-4357/ad4a54},
archivePrefix = {arXiv},
       eprint = {2311.12120},
 primaryClass = {astro-ph.GA},
       adsurl = {https://ui.adsabs.harvard.edu/abs/2024ApJ...970...95U}
}

@ARTICLE{2025A&A...693A.284K,
       author = {{Korochkin}, Alexander and {Semikoz}, Dmitri and {Tinyakov}, Peter},
        title = "{The coherent magnetic field of the Milky Way halo, the Local Bubble, and the Fan region}",
      journal = "Astron. Astrophys.",
         year = 2025,
        month = jan,
       volume = {693},
          eid = {A284},
        pages = {A284},
          doi = {10.1051/0004-6361/202451440},
archivePrefix = {arXiv},
       eprint = {2407.02148},
 primaryClass = {astro-ph.GA},
       adsurl = {https://ui.adsabs.harvard.edu/abs/2025A&A...693A.284K}
}

@ARTICLE{2025arXiv250116158K,
       author = {{Korochkin}, Alexander and {Semikoz}, Dmitri and {Tinyakov}, Peter},
        title = "{UHECR deflections in the Galactic magnetic field}",
      year={2025},
      eprint={2501.16158},
      archivePrefix={arXiv},
      primaryClass={astro-ph.HE},
      url={https://arxiv.org/abs/2501.16158}
}

@ARTICLE{2016PhRvL.116s1302P,
       author = {{Pshirkov}, M.~S. and {Tinyakov}, P.~G. and {Urban}, F.~R.},
        title = "{New Limits on Extragalactic Magnetic Fields from Rotation Measures}",
      journal = {\prl},
         year = 2016,
        month = may,
       volume = {116},
       number = {19},
          eid = {191302},
        pages = {191302},
          doi = {10.1103/PhysRevLett.116.191302},
archivePrefix = {arXiv},
       eprint = {1504.06546},
 primaryClass = {astro-ph.CO},
       adsurl = {https://ui.adsabs.harvard.edu/abs/2016PhRvL.116s1302P}
}

@ARTICLE{2024arXiv241214825N,
       author = {{Neronov}, A. and {Vazza}, F. and {Mtchedlidze}, S. and {Carretti}, E.},
        title = "{Revision of upper bound on volume-filling intergalactic magnetic fields with LOFAR}",
      journal = {arXiv e-prints},
         year = 2024,
        month = dec,
          eid = {arXiv:2412.14825},
        pages = {arXiv:2412.14825},
          doi = {10.48550/arXiv.2412.14825},
archivePrefix = {arXiv},
       eprint = {2412.14825},
 primaryClass = {astro-ph.CO},
       adsurl = {https://ui.adsabs.harvard.edu/abs/2024arXiv241214825N}
}

@ARTICLE{2021MNRAS.505.4178V,
       author = {{Vernstrom}, T. and {Heald}, G. and {Vazza}, F. and {Galvin}, T.~J. and {West}, J.~L. and {Locatelli}, N. and {Fornengo}, N. and {Pinetti}, E.},
        title = "{Discovery of magnetic fields along stacked cosmic filaments as revealed by radio and X-ray emission}",
      journal = "Mon. Not. Roy. Astron. Soc.",
         year = 2021,
        month = aug,
       volume = {505},
       number = {3},
        pages = {4178-4196},
          doi = {10.1093/mnras/stab1301},
archivePrefix = {arXiv},
       eprint = {2101.09331},
 primaryClass = {astro-ph.CO}
}

@article{Rossoni:2025vsk,
    author = {Rossoni, Simone and Sigl, G{\"u}nter},
    title = "{Anisotropy signal of ultrahigh-energy cosmic rays from a structured magnetized universe}",
    eprint = "2502.19324",
    archivePrefix = "arXiv",
    primaryClass = "astro-ph.HE",
    doi = "10.1103/4zt7-s48v",
    journal = "Phys. Rev. D",
    volume = "112",
    number = "2",
    pages = "023015",
    year = "2025"
}

@ARTICLE{2014MNRAS.440..405M,
       author = {{McCall}, Marshall L.},
        title = "{A Council of Giants}",
      journal = "Mon. Not. Roy. Astron. Soc.",
         year = 2014,
        month = may,
       volume = {440},
       number = {1},
        pages = {405-426},
          doi = {10.1093/mnras/stu199},
archivePrefix = {arXiv},
       eprint = {1403.3667},
 primaryClass = {astro-ph.GA},
       adsurl = {https://ui.adsabs.harvard.edu/abs/2014MNRAS.440..405M}
}

@article{Kalashev_2023,
    author = "Kalashev, O. and Korochkin, A. and Neronov, A. and Semikoz, D.",
    title = "{Modeling the propagation of very-high-energy \ensuremath{\gamma}-rays with the CRbeam code: Comparison with CRPropa and ELMAG codes}",
    eprint = "2201.03996",
    archivePrefix = "arXiv",
    primaryClass = "astro-ph.HE",
    doi = "10.1051/0004-6361/202243364",
    journal = "Astron. Astrophys.",
    volume = "675",
    pages = "A132",
    year = "2023"
}

@article{dolgikh2022causticlike,
    author = "Dolgikh, K. and Korochkin, A. and Rubtsov, G. and Semikoz, D. and Tkachev, I.",
    title = "{Caustic-Like Structures in UHECR Flux after Propagation in Turbulent Intergalactic Magnetic Fields}",
    eprint = "2212.01494",
    archivePrefix = "arXiv",
    primaryClass = "astro-ph.HE",
    doi = "10.1134/S1063776123060031",
    journal = "J. Exp. Theor. Phys.",
    volume = "136",
    number = "6",
    pages = "704--710",
    year = "2023"
}

@article{DOLGIKH20245295,
title = {Images of the ultra-high energy cosmic rays from point sources},
journal = {Advances in Space Research},
volume = {74},
number = {10},
pages = {5295-5301},
year = {2024},
issn = {0273-1177},
doi = {https://doi.org/10.1016/j.asr.2024.07.081},
url = {https://www.sciencedirect.com/science/article/pii/S0273117724008007},
author = {Konstantin Dolgikh and Alexander Korochkin and Grigory Rubtsov and Dmitry Semikoz and Igor Tkachev}
}

@article{AlvesBatista:2016vpy,
    author = {Alves Batista, Rafael and Dundovic, Andrej and Erdmann, Martin and Kampert, Karl-Heinz and Kuempel, Daniel and M\"uller, Gero and Sigl, Guenter and van Vliet, Arjen and Walz, David and Winchen, Tobias},
    title = "{CRPropa 3 - a Public Astrophysical Simulation Framework for Propagating Extraterrestrial Ultra-High Energy Particles}",
    eprint = "1603.07142",
    archivePrefix = "arXiv",
    primaryClass = "astro-ph.IM",
    doi = "10.1088/1475-7516/2016/05/038",
    journal = "JCAP",
    volume = "05",
    pages = "038",
    year = "2016"
}

@article{AlvesBatista:2022vem,
    author = "Alves Batista, Rafael and others",
    title = "{CRPropa 3.2 \textemdash{} an advanced framework for high-energy particle propagation in extragalactic and galactic spaces}",
    eprint = "2208.00107",
    archivePrefix = "arXiv",
    primaryClass = "astro-ph.HE",
    doi = "10.1088/1475-7516/2022/09/035",
    journal = "JCAP",
    volume = "09",
    pages = "035",
    year = "2022"
}

@article{PierreAuger:2022axr,
    author = "Abreu, Pedro and others",
    collaboration = "Pierre Auger",
    title = "{Arrival Directions of Cosmic Rays above 32 EeV from Phase One of the Pierre Auger Observatory}",
    eprint = "2206.13492",
    archivePrefix = "arXiv",
    primaryClass = "astro-ph.HE",
    reportNumber = "FERMILAB-PUB-22-491-AD-PPD-SCD-TD",
    doi = "10.3847/1538-4357/ac7d4e",
    journal = "Astrophys. J.",
    volume = "935",
    number = "2",
    pages = "170",
    year = "2022"
}

@ARTICLE{JF_GMF_1,
       author = {{Jansson}, Ronnie and {Farrar}, Glennys R.},
        title = "{A New Model of the Galactic Magnetic Field}",
      journal = {\apj},
         year = 2012,
        month = sep,
       volume = {757},
       number = {1},
          eid = {14},
        pages = {14},
          doi = {10.1088/0004-637X/757/1/14},
archivePrefix = {arXiv},
       eprint = {1204.3662},
 primaryClass = {astro-ph.GA},
       adsurl = {https://ui.adsabs.harvard.edu/abs/2012ApJ...757...14J}
}

@article{JF_GMF_2,
    author = "Jansson, Ronnie and Farrar, Glennys R.",
    title = "{The Galactic Magnetic Field}",
    eprint = "1210.7820",
    archivePrefix = "arXiv",
    primaryClass = "astro-ph.GA",
    doi = "10.1088/2041-8205/761/1/L11",
    journal = "Astrophys. J. Lett.",
    volume = "761",
    pages = "L11",
    year = "2012"
}

@article{PierreAuger:2023fcr,
    author = "Abdul Halim, Adila and others",
    collaboration = "Pierre Auger",
    title = "{An update on the arrival direction studies made with data from the Pierre Auger Observatory}",
    doi = "10.22323/1.444.0252",
    journal = "PoS",
    volume = "ICRC2023",
    pages = "252",
    year = "2023"
}

@article{Eichmann:2017iyr,
    author = {Eichmann, Bj\"orn and Rachen, J. P. and Merten, L. and van Vliet, A. and Becker Tjus, J.},
    title = "{Ultra-High-Energy Cosmic Rays from Radio Galaxies}",
    eprint = "1701.06792",
    archivePrefix = "arXiv",
    primaryClass = "astro-ph.HE",
    doi = "10.1088/1475-7516/2018/02/036",
    journal = "JCAP",
    volume = "02",
    pages = "036",
    year = "2018"
}

@article{Matthews:2018laz,
    author = "Matthews, James H. and Bell, Anthony R. and Blundell, Katherine M. and Araudo, Anabella T.",
    title = "{Fornax A, Centaurus A, and other radio galaxies as sources of ultrahigh energy cosmic rays}",
    eprint = "1805.01902",
    archivePrefix = "arXiv",
    primaryClass = "astro-ph.HE",
    doi = "10.1093/mnrasl/sly099",
    journal = "Mon. Not. Roy. Astron. Soc.",
    volume = "479",
    number = "1",
    pages = "L76--L80",
    year = "2018"
}

@article{Mollerach:2021ifa,
    author = "Mollerach, Silvia and Roulet, Esteban",
    title = "{Anisotropies of ultrahigh-energy cosmic rays in a scenario with nearby sources}",
    eprint = "2111.00560",
    archivePrefix = "arXiv",
    primaryClass = "astro-ph.HE",
    doi = "10.1103/PhysRevD.105.063001",
    journal = "Phys. Rev. D",
    volume = "105",
    number = "6",
    pages = "063001",
    year = "2022"
}

@article{Allard:2023uuk,
    author = "Allard, Denis and Aublin, Julien and Baret, Bruny and Parizot, Etienne",
    title = "{What can be learnt from UHECR anisotropies observations - II. Intermediate-scale anisotropies}",
    eprint = "2305.17811",
    archivePrefix = "arXiv",
    primaryClass = "astro-ph.HE",
    doi = "10.1051/0004-6361/202347034",
    journal = "Astron. Astrophys.",
    volume = "686",
    pages = "A292",
    year = "2024"
}

@ARTICLE{Giacalone_1999,
       author = {{Giacalone}, J. and {Jokipii}, J.~R.},
        title = "{The Transport of Cosmic Rays across a Turbulent Magnetic Field}",
      journal = {\apj},
         year = 1999,
        month = jul,
       volume = {520},
       number = {1},
        pages = {204-214},
          doi = {10.1086/307452},
       adsurl = {https://ui.adsabs.harvard.edu/abs/1999ApJ...520..204G}
}

@ARTICLE{2021JCAP...04..065K,
       author = {{Kuznetsov}, M. Yu. and {Tinyakov}, P.~G.},
        title = "{UHECR mass composition at highest energies from anisotropy of their arrival directions}",
      journal = "JCAP",
         year = 2021,
        month = apr,
       volume = {2021},
       number = {4},
          eid = {065},
        pages = {065},
          doi = {10.1088/1475-7516/2021/04/065},
archivePrefix = {arXiv},
       eprint = {2011.11590},
 primaryClass = {astro-ph.HE},
       adsurl = {https://ui.adsabs.harvard.edu/abs/2021JCAP...04..065K}
}

@ARTICLE{2024PhRvL.133d1001A,
       author = {{Telescope Array Collaboration}, {Abbasi}, R.~U. and {Abe}, Y. and {Abu-Zayyad}, T. and others },
        title = "{Isotropy of Cosmic Rays beyond {}10$^{20}$ eV Favors Their Heavy Mass Composition}",
      journal = {\prl},
         year = 2024,
        month = jul,
       volume = {133},
       number = {4},
          eid = {041001},
        pages = {041001},
          doi = {10.1103/PhysRevLett.133.041001},
archivePrefix = {arXiv},
       eprint = {2406.19287},
 primaryClass = {astro-ph.HE},
       adsurl = {https://ui.adsabs.harvard.edu/abs/2024PhRvL.133d1001A}
}

@ARTICLE{2024PhRvD.110b2006A,
       author = {{Telescope Array Collaboration}, {Abbasi}, R.~U. and {Abe}, Y. and {Abu-Zayyad} T. and others},
        title = "{Mass composition of ultrahigh energy cosmic rays from distribution of their arrival directions with the Telescope Array}",
      journal = {\prd},
         year = 2024,
        month = jul,
       volume = {110},
       number = {2},
          eid = {022006},
        pages = {022006},
          doi = {10.1103/PhysRevD.110.022006},
archivePrefix = {arXiv},
       eprint = {2406.19286},
 primaryClass = {astro-ph.HE},
       adsurl = {https://ui.adsabs.harvard.edu/abs/2024PhRvD.110b2006A}
}

@ARTICLE{1996ApJ...472L..89W,
       author = {{Waxman}, Eli and {Miralda-Escude}, Jordi},
        title = "{Images of Bursting Sources of High-Energy Cosmic Rays: Effects of Magnetic Fields}",
      journal = "Astrophys. J. Lett.",
         year = 1996,
        month = dec,
       volume = {472},
        pages = {L89},
          doi = {10.1086/310367},
archivePrefix = {arXiv},
       eprint = {astro-ph/9607059},
 primaryClass = {astro-ph},
       adsurl = {https://ui.adsabs.harvard.edu/abs/1996ApJ...472L..89W}
}

@ARTICLE{2016PhRvD..93f3002H,
       author = {{Harari}, Diego and {Mollerach}, Silvia and {Roulet}, Esteban},
        title = "{Angular distribution of cosmic rays from an individual source in a turbulent magnetic field}",
      journal = {\prd},
         year = 2016,
        month = mar,
       volume = {93},
       number = {6},
          eid = {063002},
        pages = {063002},
          doi = {10.1103/PhysRevD.93.063002},
archivePrefix = {arXiv},
       eprint = {1512.08289},
 primaryClass = {astro-ph.HE},
       adsurl = {https://ui.adsabs.harvard.edu/abs/2016PhRvD..93f3002H}
}

@article{Kim:2025ykm,
    author = "Kim, Jihyun and Ivanov, Dmitri and Kawata, Kazumasa and Sagawa, Hiroyuki and Thomson, Gordon",
    collaboration = "Telescope Array",
    title = "{Telescope Array Surface Detector Medium-scale Anisotropy Analyses}",
    doi = "10.22323/1.484.0097",
    journal = "PoS",
    volume = "UHECR2024",
    pages = "097",
    year = "2025"
}

@article{Taylor:2023qdy,
    author = "Taylor, A. M. and Matthews, J. H. and Bell, A. R.",
    title = "{UHECR echoes from the Council of Giants}",
    eprint = "2302.06489",
    archivePrefix = "arXiv",
    primaryClass = "astro-ph.HE",
    doi = "10.1093/mnras/stad1716",
    journal = "Mon. Not. Roy. Astron. Soc.",
    volume = "524",
    number = "1",
    pages = "631--642",
    year = "2023"
}

@article{PierreAuger:2023gmj,
    author = "Abdul Halim, Adila and others",
    collaboration = "Pierre Auger",
    title = "{Mass Composition from 3 EeV to 100 EeV using the Depth of the Maximum of Air-Shower Profiles Estimated with Deep Learning using Surface Detector Data of the Pierre Auger Observatory}",
    doi = "10.22323/1.444.0278",
    journal = "PoS",
    volume = "ICRC2023",
    pages = "278",
    year = "2023"
}

@article{Bister:2023icg,
    author = "Bister, Teresa and Farrar, Glennys R.",
    title = "{Constraints on UHECR Sources and Extragalactic Magnetic Fields from Directional Anisotropies}",
    eprint = "2312.02645",
    archivePrefix = "arXiv",
    primaryClass = "astro-ph.HE",
    doi = "10.3847/1538-4357/ad2f3f",
    journal = "Astrophys. J.",
    volume = "966",
    number = "1",
    pages = "71",
    year = "2024"
}

@article{Farrar:2012gm,
    author = "Farrar, Glennys R. and Jansson, Ronnie and Feain, Ilana J. and Gaensler, B. M.",
    title = "{Galactic magnetic deflections and Centaurus A as a UHECR source}",
    eprint = "1211.7086",
    archivePrefix = "arXiv",
    primaryClass = "astro-ph.HE",
    doi = "10.1088/1475-7516/2013/01/023",
    journal = "JCAP",
    volume = "01",
    pages = "023",
    year = "2013"
}

@article{Keivani:2014kua,
    author = "Keivani, Azadeh and Farrar, Glennys R. and Sutherland, Michael",
    title = "{Magnetic Deflections of Ultra-High Energy Cosmic Rays from Centaurus A}",
    eprint = "1406.5249",
    archivePrefix = "arXiv",
    primaryClass = "astro-ph.HE",
    doi = "10.1016/j.astropartphys.2014.07.001",
    journal = "Astropart. Phys.",
    volume = "61",
    pages = "47--55",
    year = "2014"
}

@article{Farrar:2017lhm,
    author = "Farrar, Glennys R. and Sutherland, Michael S.",
    title = "{Deflections of UHECRs in the Galactic magnetic field}",
    eprint = "1711.02730",
    archivePrefix = "arXiv",
    primaryClass = "astro-ph.HE",
    doi = "10.1088/1475-7516/2019/05/004",
    journal = "JCAP",
    volume = "05",
    pages = "004",
    year = "2019"
}
\end{document}